# Preparation of Cobalt Thin Films by Sputtering Systems and Its Magnetic Characterization

Mustafa Erkovan [a]

[a] Gebze Institute of Technology, Department of Physics, 41400 Gebze/KOCAELİ, TURKEY

**Abstract**

Different thicknesses of cobalt thin films were growth by magnetron sputtering deposition techniques. The films thicknesses were determined with X ray Photoelectron Spectroscopy (XPS) and Quartz Crystal Monitoring (QCM). XPS is also used to determinate the films quality. The films magnetic properties were determinated by Ferromagnetic Resonance (FMR) technique.

[*] Corresponding author. Tel.: +90 262 605 17 88; fax: +90 (262) 653 84 90; e-mail: merkovan@gyte.edu.tr

# 1. Introduction

Cobalt is used in the preparation of wear-resistant, and high-strength alloys and cobalt is widely used in magnetic recording because it is the only one of the three room temperature ferromagnets that has unaxial symmetry and therefore can be used in digital recording. Cobalt has a relative permeability two thirds that of iron. Metallic cobalt occurs as two crystallographic structures hcp and fcc. Cobalt based superalloys consume most of the produced cobalt. The temperature stability of these alloys makes them suitable for use in turbine blades for gas turbines and jet aircraft engines, though nickel-based single crystal alloys surpass them in this regard. Cobalt based alloys are also corrosion and wear-resistant [1]. Special cobalt chromium molybdenum alloys are used for prosthetic parts such as hip and knee replacements [2]. Cobalt alloys are also used for dental prosthetics, where they are useful to avoid allergies to nickel [3]. Some high speed steels also use cobalt to increase heat and wear-resistance. The special alloys of aluminium, nickel, cobalt and iron, known as Alnico, and of samarium and cobalt (samarium-cobalt magnet) are used in permanent magnets [4]. Lithium cobalt oxide ($LiCoO_2$) is widely used in Lithium ion battery electrodes [5]. Nickel-cadmium (NiCd) and nickel metal hydride (NiMH) batteries also contain significant amounts of cobalt. Several cobalt compounds are used in chemical reactions as catalysts. Cobalt acetate is used for the production of terephthalic acid as well as dimethyl terephthalic acid, which are key compounds in the production of Polyethylene terephthalate. The steam reforming and hydrodesulfuration for the production of petroleum, which uses mixed cobalt molybdenum aluminium oxides as a catalyst, is another important application [5]. Cobalt and its compounds, especially cobalt carboxylates (known as cobalt soaps), are good oxidation catalysts. They are used in paints, varnishes, and inks as drying agents through the oxidation of certain compounds [5]. The same carboxylates are used to improve the adhesion of the steel to rubber in steel-belted radial tires [5].

## 2. Experimental

XPS analyses were performed using SPECS-Phoibus 150 Electron Analyzer (with Medium Magnification, 3400eV detector voltage for 9 Channels, and 5mm slid) and SPECS-XR50 x-ray source (with 350W power).

Magnetic measurements were performed using Bruker EMX model EPR spectrometer.

## 3. Results

### 3.1. Sample Preparation

Co films were grown on $SiO_2$ (100) substrate by rf-magnetron deposition technique attached to a cluster ultrahigh vacuum chamber, containing analytical chamber besides to sputtering chamber and a sample-loading section. Oriented and polished silicon substrates were obtained commercially MaTech. Prior to insertion into the sample loading section, the substrates were washed with methanol and ethanol in sonic cleaner for 5 minutes. The substrates were transferred under the etching gun to clean by cycles of $Ar^+$ sputtering followed by annealing at ~ 600° C for 30 minutes. The power of magnetron rf-sputtering gun was 70W for the 300 second etching process. During the etching process, Ar pressure was maintained at $2x10^{-3}$ Torr. After finishing sample preparation step, in UHV condition the silicon substrates were transferred under the rf-magnetron gun in the sample preparing chamber. Rf-magnetron gun attached Co-target is operated at 24 W. The distance between the target and the substrate is 50 mm. Although the base pressure of the sample preparing chamber is $5 \times 10^{-9}$ mbar, high purity Ar (99,9999%) is leaked to the chamber with 2 sscm flow rate so that the pressure is fixed to $2.6 \times 10^{-3}$ mbar during the period of sputtering deposition process. MKS flow meter 1179A was used to control gas flow with ±1% sensitivity. The deposition rate is monitored with the quartz crystal thickness monitor (QCM), calibrated by using XPS. During the XPS data acquisition the sample current was about 2μA. Although $Si2p_{3/2}$ is the most favorable photoemission peak for thickness calibration using XPS signals, the attenuation of $Si2p_{1/2}$ peak coming from the surface of the film is observed as a function of cobalt exposure, shown in Figure 1a. Since Si2p$_{3/2}$ and Co2s photoemission peaks sit close at the spectroscopy it makes peak-fitting-process hard to calculate Co converges.

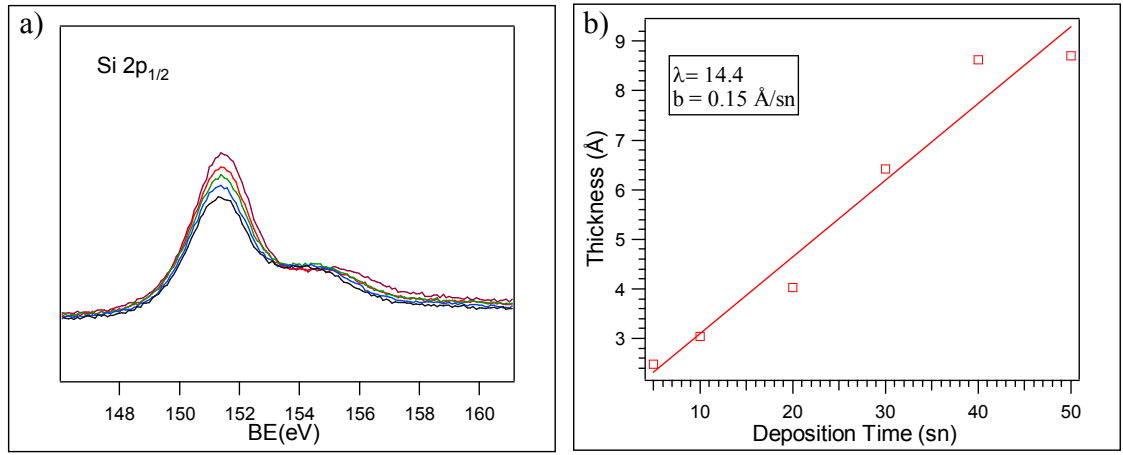

**Figure 1.** Since the peaks of Co2s and Si2p$_{3/2}$ are at the same place at the photoemmision spectrum the Si2p$_{1/2}$ attenuation was used to calculate Co coverage for this work. The **figure 1a** shows the attenuation of the Si2p$_{1/2}$ photoemission intensity as a function of Co film thickness. The Co film thickness is calculated from equation 1 using a value of 14.4 Å for the Si2p$_{1/2}$ inelastic mean free path. **Figure 1b** shows thickness versus deposition time.

In converting ratios to Co thickness, the electron mean free path is calculated by using the TPP formula1. The relationship between the Co film thickness, x, and the Si2p$_{1/2}$ photoemission attenuation can be written as [6]

$$\ln\left(\frac{I_s(x)}{I_s(0)}\right) = -\frac{x}{\lambda_s}, \quad (1)$$

where $I_s(x)$ is the substrate photoemission intensity when covered by a Cr film of thickness x, $I_s(0)$ is the clean substrate photoemission intensity, and $\lambda_s$ is the inelastic mean free path of the substrate photoelectrons [6]. Electrons with kinetic energy of ≈ 151 eV (AlK$_\alpha$ excitation) for Si 2p$_{1/2}$ have an inelastic mean free path of approximately 14.4 Å. Figure 1b shows thickness versus deposition time. The deposition rate is found 0.15 Å from the slope of the line in Figure 1b. Base on this calibration, the various thicknesses of Co films (500 Å, 1000 Å and 1500 Å) were prepared. After every deposition, XPS with AlK$_\alpha$ radiation (hν=1486.6 eV) provided twin anode Specs X-ray Source XR 50 is used to confirm the sample cleanliness and composition.

## 3.2. FMR Measurements

All spectra were obtained at microwave frequencies X-band (9.8 GHz). We note that all FMR spectra are presented in first derivative format since an AC modulation magnetic field of 100 kHz was applied parallel to the external DC magnetic field to get field derivative curves of microwave resonant absorption. The external magnetic field was swept in the range 0-20 kG in the horizontal direction. Data were collected from each film for two different orientations at room temperature. The first one is in plane (in-plane geometry, $\theta_H = 90^0$) and the second one is perpendicular to the film plane (out of- plane $\theta_H = 0^0$). In order to rotate samples about vertical direction in the horizontal DC magnetic filed a goniometer was used. The spectra have been recorded as a function of angle between DC magnetic field and the film normal, it is shown in Figure 2 [7].

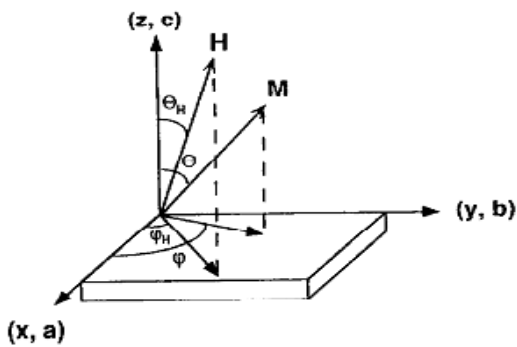

**Figure 2.** The axis system used in FMR measurement. Here H is the external magnetic field and M is the magnetization vector. The microwave component of the magnetic filed is always along the x-axis while the DC magnetic filed remains in the yz plane which is also perpendicular to the rotation axis.

Figure 3 shows room temperature FMR spectra taken at X-band for each Co film thicknesses for comparison. The three spectra with resonance peaks at lower fields represent in-plane geometry while at higher fields represent out of-plane where magnetic field is perpendicular to the film plane. Resonance lines for parallel geometries of all the films are shifted to very low fields. The large differences between the resonance fields for parallel and perpendicular geometries are related to the demagnetizing field. The value of this separation of the resonance peaks is slightly changed with film thickness. For thicker film the separation is slightly larger. These values are 18.5 kG, 19.5 kG and 20.5 kG respectively for 500 Å, 1000 Å and 1500 Å film thicknesses.

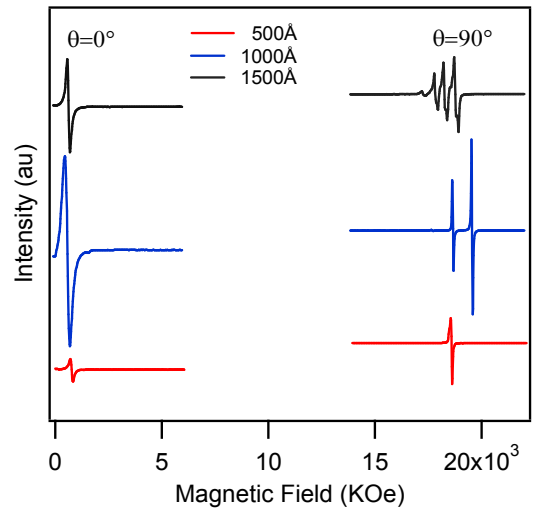

**Figure 3.** Room temperature experimental SWR spectra taken X-band from three different cobalt thicknesses for both parallel $\theta = 90^0$ and perpendicular $\theta = 0^0$ geometries.

The other most noticeable feature in Figure 3 is the number of peaks in the spectra. The cobalt film with 500 Å thickness there is only one main peak for both parallel and perpendicular geometry. However a close looking to the spectrum reveals a very weak mode at the lower side of main peak. The cobalt film with 1000 Å thickness has two strong peaks which are well resolved at perpendicular position. Relative intensities of two modes in perpendicular geometry are comparable. The peak at higher magnetic filed will be named first mode and the peak is at lower field will be named second mode. A very weak peak also appeared at the lower side of the second mode. For some selected angles angular dependence of SWR spectra for this film (1000 Å) are given in Figure 4. As the magnetic field is rotated away from film normal towards the film plane such as 2.5 degree intensity of main mode improved while intensity of second mode becomes much weaker. The resonance peaks of both first and second modes shift to lower fields and separation between first and second mode slightly increased. The second mode and third mode are disappeared at $3.5^0$ away from film normal and only observable peak was the main mode. The line widths for both modes are quite increased as the magnetic field is rotated away from film normal. All these features indicate that this film has easy plane and relatively weak surface anisotropy energy.

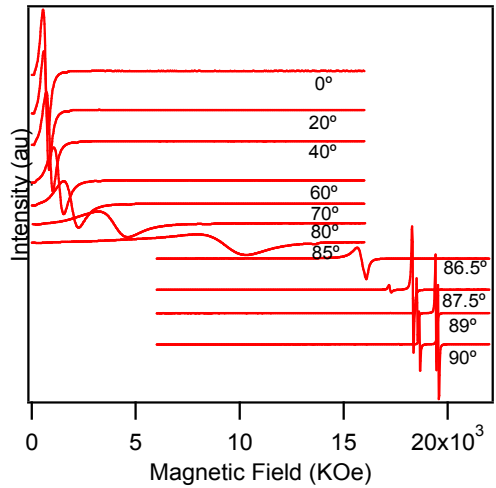

Figure 4. Angular behavior of experimental SWR spectra taken at room temperature from 1000 Å thickness Co thin film. Magnetic filed is parallel to film surface where the spectrum corresponding $0^0$ degree and it is perpendicular to the film surface where angle is $90^0$.

On the other hand, the number of observed peaks is larger for the thickest film, 1500 Å (shown in figure 3). As we can see clearly there are many modes in the spectra taken at perpendicular geometry. As magnetic field rotated away from film normal the number of peaks decreases and finally all modes except the main mode are disappeared at $4^0$ away from film normal. The line separation between modes is smaller compare to 1000 Å thickness. This separation between bulk modes are expected to be inversely proportional to the square of the film thickness [2]. The line width of spectra taken for 1500 Å is little broader compare to 1000 Å cobalt film resulting in overlaps. However qualitative behavior of the spectra is similar to that of the 1000 Å thick film.

Figure 5 shows the angular dependence of spectra for some selected angles for 1500 Å. Here also the line widths for all modes increase as the magnetic field is rotated away from film normal and takes its maximum value at the angle $5^0$ and line width becomes narrow again when magnetic filed is parallel to film surface. This behavior was same for thinner films. It should be noted that in Figure 3 the single mode of the thinnest film corresponds to the second mode in the spectrum for thicker film while it corresponds to the first mode at the highest field of the spectrum for the thickest film.

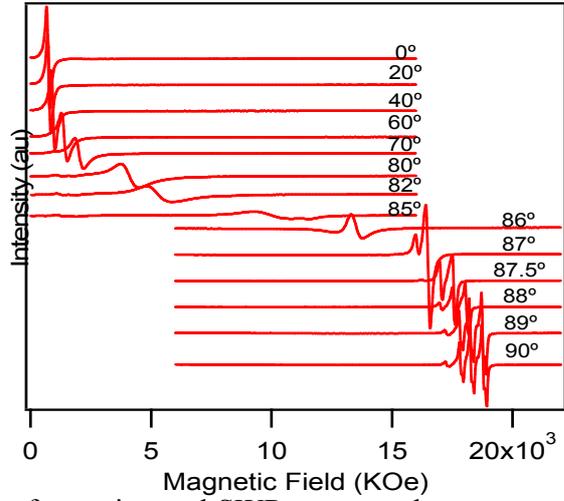

**Figure 5.** Angular behavior of experimental SWR spectra taken at room temperature from 1500 Å thickness Co thin film. Direction for magnetic filed is same as in the figure 4.

In figures 6 the angular dependence of the resonance fields (out of plane position) for each samples are presented. As it can seen from this angular dependence is highly anisotropic for each film. The resonance filed values for each mode are maximum at $90^0$ where magnetic field is perpendicular to the film surfaces. Figure 6 also shows the angular dependence of other modes in the spectra.

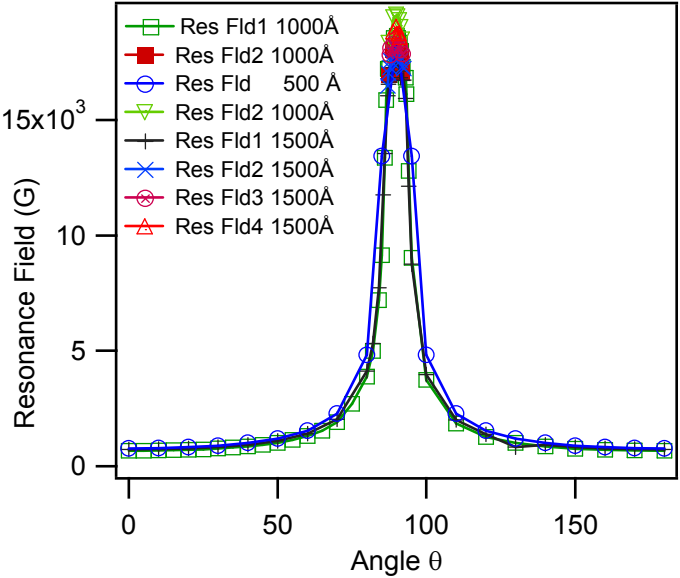

**Figure 6.** The angular dependence of the resonance fields for all samples taken at room temperature for the perpendicular geometry. The value of resonance filed is maximum at $90^0$ where magnetic field is perpendicular to the film surfaces.

DC magnetic field was also rotated in the film plane for each cobalt films. Noticeable anisotropy was observed. The angular dependence of resonance filed values is presented in the figure 7. As it can bee seen angular dependency of resonance lines for thicker film is more anisotropic than thinner film. The difference between maximum and minimum resonance value is nearly 60 G. The directions of the easy and hard axes of magnetization were obtained from in plane rotation. The easy axis corresponds to the minimal in-plane resonance filed, and the hard axis corresponds to the maximal in-plane resonance filed.

### 3.3. Theorical Model

The theoretical energy terms in FMR are calculated by using the following expression.

$$E = \mathbf{M}.\mathbf{H} + K_{eff} \cos^2(\theta) + K_1 \sin^2(\theta) + K_2 \sin^4(\theta) \qquad (2)$$

where $\mathbf{M}$ is the magnetization, $\mathbf{H}$ the external DC magnetic field, $K_{eff}(2\pi M^2 \; K_b)$ the effective unaxial volume (bulk) anisotropy parameter. Here $K_1$ and $K_2$ are, respectively, the unaxial and biaxial surface anisotropy energy parameters [7-13] and $\theta$ is angle of $\mathbf{M}$ with respect to the film normal, as shown fig. 2 [7].

For any arbitrary direction of static field $\mathbf{H}$, the basic dispersion relation for spin waves is given [13] as,

$$\left(\frac{w}{\gamma}\right)^2 = \left(\frac{1}{M_s \sin^2\theta}\frac{\partial^2 E}{\partial \varphi^2} + Dk_n^2\right) \times \left(\frac{1}{M_s}\frac{\partial^2 E}{\partial \theta^2} + Dk_n^2\right) \quad \left(\frac{1}{M_s \sin\theta}\frac{\partial^2 E}{\partial \varphi \partial \theta}\right)^2 \qquad (3)$$

where $\gamma$ is the gyro-magnetic ratio, $w$ is the microwave frequency, $\theta$ and $\varphi$ are the usual spherical polar angles for $\mathbf{M}$, $D(=2A/M_s)$ is the exchange parameter and $k_n$ is the spin wave vector for nth mode. Here $\theta$ is determined by the static equilibrium condition for the magnetization

$$\mathbf{M.H}\,Sin(\theta - \theta_H) + K_{eff}\,Sin(2\theta) = 0 \qquad (4)$$

Thus, using Eqs. (3) and Eqs. (4) in the general resonance condition which given by Eqs. (2) one can obtain the following expression for resonance field for nth SWR mode;

$$[\frac{w}{\gamma}]^2 = [H\frac{sin(\theta_H)}{sin(\theta)} + Dk_n^2] \times [HCos(\theta_H - \theta) + (2K_{eff}/M_s)Cos^2\theta + Dk_n^2] \qquad (5)$$

for any direction $\theta_H$ of the measurement field. The real and imaginary parts of spin wave vector $k_n$ are determined from the following expression [11,12]

$$tan(k_n L) = \frac{k_n(P_1 + P_2)}{k_n^2 - P_1 P_2} \qquad (6)$$

$$tanh(k_n L) = -\frac{k_n(P_1 + P_2)}{k_n^2 + P_1 P_2} \qquad (7)$$

for the bulk and the surface modes, respectively. Here $L$ is the film thickness and $P_i$'s are the surface pinning parameters which are defined for either surface as [10,11].

$$P = -\frac{2(K_1 + K_2)L}{A}Cos(2\theta) + \frac{2K_2 L}{A}Cos(4\theta) \qquad (8)$$

The second term of Eqs. (8) is the normal derivative of the saturation magnetization and the first one comes from the surface energy the analytical form of the absorption curve is more complicated and has been given in a previous work. [10]

## 4. Discussion

After collecting enough spectra (film was rotated relative to magnetic field for each $2.5^0$ degree) for each film at room temperature the theoretical model that explained above has been used for analyzing the data. The parameters are found from these theoretical simulations are presented in the Table 1. The calculated spectra are presented in figure 6 along with corresponding experimental ones. As seen in the figure 6 fairly good agreements are obtained between experimental and theoretical spectra as a function of the applied filed angle. The symbol in the figure represents the values which were obtained from theoretical model for the calculated resonance field to the corresponding experimental ones (solid lines). It should be emphasized that the fitting parameters were first optimized for perpendicular position for each different thickness films and then all spectra with different angles were fitted with these fixed parameters. A constant value of 3250 Gauss was used for $\omega/\gamma$ which corresponds to a g-value of 2.16 for cobalt. This value was same for all three films.

From analysis of angular dependency of the SWR spectra for 500 Å, the strong peak at high field value for perpendicular geometry was identified as the surface mode. The weak peak at the lower side of strong mode was identified as the bulk mode. Same analyses were carried for both 1000 Å and 1500 Å cobalt films. For thinner film, the strong peak at higher field value and relatively strong peak at lower filed for perpendicular geometry were identified as the surface mode and first bulk mode respectively. Relatively weak peak appeared at the lower side of first bulk mode was also identified as second bulk mode. Sharp single surface mode indicates that the boundary conditions on both surfaces are the exactly same. The pinning parameters are governed mainly by surface anisotropy energy including uniaxial term in Eq. 2. The surface anisotropy has an easy-plane character. For the thickest film the two modes at the highest filed are identified as first and second surface modes. Their anisotropy parameters are quite different indicating that boundary conditions on both surfaces are different and are affected by the film-substrate interface for this thickness. Similar to 1000 Å film uniaxial surface anisotropy has an easy-plane character.

As shown in Table 1. Spin wave exchange stiffness constant, D (=2A/M$_s$) was same from all of the as-prepared 500, 1000 and 1500 Å cobalt films at room temperature. This value is in between the values given in literature [14,15]. The effective volume (bulk) anisotropy parameter, K$_{eff}$, which stands for demagnetization field ($-4\pi M_s$) is increased from 500 Å film to 1000 Å film but the value decreased significantly for 1500 Å film. On the other hand high anisotropy was observed for in plane the measurements compare to 1000 Å thick film. So the reason for decrease in K$_{eff}$ value is connected with in-plane anisotropy. The uniaxial surface anisotropy parameters for both surfaces, K$_1^{s1}$ (free surface) and K$_1^{s2}$ (substrate surface) were varied with thickness and were noticeably increased for 1500 Å film especially for the free side of the film. While the symmetric magnetic boundary conditions were found for both 500, 1000 Å cobalt films, asymmetric magnetic boundary condition were found for 1500 Å since both surface modes are clearly resolved and have quite different values. A difference in pinning conditions at the two surfaces are the reason for two surface modes.

| Samples | D (cm$^2$ G) | 2K$_{eff}$/ M (G) | K$_1^{s1}$ L / A | K$_1^{s2}$ L / A | ΔH (G) |
|---|---|---|---|---|---|
| 500 Å | 0.9 x 10$^{-9}$ | 15250 | -6.10 | -6.10 | 100 |
| 1000 Å | 0.9 x 10$^{-9}$ | 15575 | -9.41 | -9.41 | 100 |
| 1500 Å | 0.9 x 10$^{-9}$ | 14605 | -15.10 | -9.50 | 110 |

**Table 1.** Spin wave exchange stiffness constant.

## 5. Conclusion

Different thicknesses of cobalt thin films were very well growth by magnetron sputtering deposition techniques. The films thicknesses were determined with X ray Photoelectron Spectroscopy (XPS) and Quartz Crystal Monitoring (QCM). XPS results were shown high quality of Co films. The films magnetic properties were determined by Ferromagnetic Resonance (FMR) technique and experimental results were shown very good compatible with theoretical modeling.

**Acknowledgments**

The authors also express their special thanks to Dr. Osman Öztürk, Dr. Bekir Aktaş, Dr. Mustafa Özdemir and Dr. Sibel T Öztürk because of their help.